\documentstyle[b98proc,epsfig]{article}
\include{epsfig}

\def\Journal#1#2#3#4{{#1} {\bf #2}, #3 (#4)}

\def\PRD{{\em Phys. Rev.} D}

\begin{document}
\title{
N DELTA - TRANSITION FORM FACTORS AT LOW MOMENTUM TRANSFER
}

\author{G. C. GELLAS, C. N. KTORIDES, G. I. POULIS}
\address{Univ. of Athens, Dept. of Physics, Athens, Greece}
\author{T. R. HEMMERT}
\address{FZ J\"ulich, IKP (Theorie), J\"ulich, Germany}

\maketitle

\abstracts{
The three complex form factors entering
the $\Delta\rightarrow N\gamma^\ast$ vertex
are calculated to ${\cal O}(\varepsilon^3)$ in the framework of a
chiral effective theory with explicit $\Delta$(1232) degrees of freedom.
Furthermore, the role of presently unknown low energy constants that affect the
values
of EMR and CMR is elucidated.
}

\section {Introduction}

In this talk we present a new calculation \cite{Our} of the
$\Delta N$ transition form factors at low momentum
transfer  $t\equiv q^2$ in an effective chiral lagrangian framework. For a
survey of the ongoing experimental efforts regarding this
fundamental transition of the nucleon to its first excited state we
refer
to the contributions of P. Bartsch, M.O. Distler and R.W. Gothe in these
proceedings. In our calculation a scale $\varepsilon =
\{p,m_\pi,\delta\}$ denoting, collectively, small momenta, the
pion mass and the Delta-nucleon mass splitting
 is used to establish a systematic power-counting, thus telling us
precisely which diagrams/vertices have to be included if we want to calculate
up
to a certain order in $\epsilon$.
This approach allows for an efficient inclusion of $\Delta$(1232) degrees of
freedom
consistent with the underlying chiral symmetry of QCD and is referred to as the
``Small Scale Expansion'' (SSE) \cite{long}, constituting a phenomenological
extension of Heavy Baryon Chiral Perturbation Theory \cite{HB}. So far, we
have performed the calculation to ${\cal O}(\varepsilon^3)$ and in the
process
have also established the appropriate relations between the
non-relativistic microscopic
calculation as well as phenomenological parameterizations of the  
transition current.
We find that
the $q^2$ evolution of the three complex form factors is completely determined
by the dynamics of the nucleon's pion cloud governed by chiral symmetry.
Finally, utilizing previous calculations performed
at $q^2=0$ we discuss the role of relevant low energy parameters for the sought
after
multipole ratios EMR$(q^2)$ and CMR$(q^2)$.

\section{Non-relativistic Reduction of the $\Delta\rightarrow N\gamma^\ast$
Vertex}

Demanding Lorentz covariance, gauge invariance and parity conservation the
most general form of the $\Delta\rightarrow N\gamma^\ast$ radiative decay
amplitude is described by three form factors $G_i(q^2)$, i=1,2,3,
\begin{eqnarray}\label{eq1}
i{\cal M}_{\Delta N\gamma}^{full}
&=&\frac{e}{2M_N} \bar{u}(p) \gamma_5 \left[
   G_1(q^2)( \not{q} \epsilon_\mu -  \not{\epsilon} q_\mu )
 + {G_2(q^2)\over 2M_N} (p\cdot \epsilon\, q_\mu - p\cdot q\,
\epsilon_\mu) \right.\nonumber \\
& &\phantom{\frac{e}{2M_N} \bar{u}(p) \gamma_5 }
\left. + {G_3(q^2)\over 2 (M_\Delta-M_N)} (q\cdot \epsilon\, q_\mu - q^2
\epsilon_\mu)\right]u^\mu_\Delta(p_\Delta) \, .
\end{eqnarray}

Here $M_N\,(M_\Delta)$ is the nucleon (Delta) mass, $p^\mu ,p_{\Delta}^\mu$
denotes the
four-momentum of the
nucleon, Delta and $q^\mu$, $\epsilon^\mu$ represent the photon four-momentum
and
polarization vectors, respectively.
A SSE calculation of the radiative
vertex to ${\cal O}(\varepsilon^3)$ entails a restriction to ${\cal
O}(1/M_N^2)$ accuracy \cite{long}. We must, therefore, compute the most
general form of Eq.(\ref{eq1}) consistent with this estimate. Our
calculation will be performed in the $\Delta(1232)$ rest frame which is
convenient for the identification of various multipoles \cite{Our}. In
this frame, taking into account that
to ${\cal O}(\varepsilon^3)$ the photon energy obeys $\omega$ = $\delta$ +
${\cal O}(1/M_N)$ and assuming that each form factor scales at least as 
${\cal O}(M_N^0)$ one determines
\begin{eqnarray}\label{exp2}
i{\cal M}&=&e \ \bar{u}_v(r_N)\left\{
 (S\cdot \epsilon) q_\mu
\left[ \frac{G_1(q^2)}{M_N}+{\cal O}(1/M_N^3)  \right] \right.
\nonumber \\
& &+ (S\cdot q) \epsilon_\mu
\left[ -\frac{G_1(q^2)}{M_N}-\frac{\delta G_1(0)}{2M_N^2}+\frac{\delta
G_2(q^2)}{4M_N^2}
+\frac{q^2G_3(q^2)}{4M_N^2\delta}+... \right] \nonumber \\
& &+ (S\cdot q) (v\cdot\epsilon) q_\mu
\left[ \frac{G_1(0)}{2M_N^2} - \frac{G_2(q^2)}{4M_N^2}
+{\cal O}(1/M_N^3) \right] \nonumber \\
& &\left. + (S\cdot q) (q \cdot \epsilon) q_\mu
\left[ -\frac{G_3(q^2)}{4M_N^2\delta}
+ {\cal O}(1/M_N^3) \right]
\right\} u_{v,\Delta}^{\mu,i=3}(0)\ ,
\end{eqnarray}
where $v_\mu$ denotes the rest frame four velocity of the Delta field and
$S_\mu$
corresponds
to the Pauli-Lubanski vector. From Eq.(\ref{exp2}) one can already see that to
${\cal O}(\epsilon^3)$ one is sensitive to the first two orders in the chiral
expansion
of $G_1$, whereas the quadrupole form factors $G_2,G_3$ only start at
${\cal O}(\epsilon^3)$ and therefore only their leading behavior is determined
in this
calculation. Note that we had to introduce a particular mass dependence
proportional to
$\delta^{-1}$ accompanying the form factor $G_3(q^2)$ in Eq.(\ref{eq1}) in
order to
achieve consistency with the chiral ${\cal O}(\epsilon^3)$ calculation
\cite{Our}.

\section{The Calculation}

Independent of any choice
of gauge one finds that even for finite $q^2$ only two
one-loop diagrams---the well known
$\Delta\rightarrow N\gamma$ triangle diagrams \cite{Butler} corresponding
to the photon been attached
 to two pions and the intermediate baryon state being a nucleon and a  
Delta,
respectively---contribute to this order. Due to the fact that the
$\Delta\rightarrow N\gamma^\ast$ transition starts with a magnetic dipole
amplitude M1,
one finds no ${\cal O}(\varepsilon)$ vertex but has to take into
account
all ${\cal O}(\varepsilon^2)$ and ${\cal O}(\varepsilon^3)$ tree contributions.
For the Born contributions one obtains \cite{Our}
\begin{eqnarray}
i{\cal M}_{\Delta\rightarrow
N\gamma^\ast}^{Born}&=&e\;\bar{u}_v(r_N)\left[
S\cdot\epsilon\;q_\mu\left(\frac{-b_1}{2M_N}+\frac{(2E_1-D_1)\delta}{4
M_{N}^2}
\right)\right.\nonumber \\
& &+S\cdot q\; \epsilon_\mu\left(\frac{b_1}{2M_N}-\frac{E_1\delta}{2
M_{N}^2}+
\frac{(b_1+2b_6) \;\delta}{4 M_{N}^2}\right)
\nonumber \\
& &\left.+\epsilon_0\;S\cdot q\;q_\mu\left(\frac{D_1}{4M_{N}^2}
-\frac{b_1+2b_6}
{4 M_{N}^2}\right)\right]u_{v,\Delta}^{\mu,i=3}(0),
\label{eq:born}
\end{eqnarray}
where $b_1, b_6$ correspond to the two leading ({\it i.e.} ${\cal
O}(\varepsilon^2)$)
$\Delta N\gamma$ couplings \cite{Davidson}, whereas $D_1,E_1$ are new
\cite{Our}
$\Delta N\gamma$ couplings of ${\cal O}(\varepsilon^3)$, which also take part
in the
renormalization of the loop diagrams.

The loop contributions can be formally written as \cite{Our}
\begin{eqnarray}\label{HBXPT}
i{\cal M}_{\Delta\rightarrow
N\gamma^\ast}^{Loop}&=&{2 e g_{\pi N\Delta}\over F_\pi^2} \ \bar{u}_v(r_N)
\Bigl\{
  (S\cdot \epsilon) q_\mu
\left[ g_A F_N(t) + \chi g_1 F_\Delta(t) \right]
        \nonumber \\
& & +(S\cdot q) \epsilon_\mu
\left[ g_A G_N(t) + \chi g_1 G_\Delta(t) \right]+(S\cdot q) q_\mu
\Bigl( \epsilon_0 [ g_A J_N(t)
      \Bigr.\nonumber \\
& &\Bigl. + \chi g_1 J_\Delta(t) ]
     (q\cdot\epsilon)\left[ g_A H_N(t) + \chi g_1 H_\Delta(t) \right]
    \Bigr) \Bigr\} u_{v,\Delta}^{\mu\,i=3}(0) \, ,
\end{eqnarray}
where the functions $I_{N,\Delta}$, $I\in\{F,G,J,H\}$ are given in terms of
integrals over a
Feynman parameter. The last structure proportional
to $q\cdot \epsilon$ is the one which cannot be reproduced by 
Eq.(\ref{exp2})
without the particular mass term accompanying the $G_3$ form factor in
Eq.(\ref{eq1})
as discussed
above. Comparing Eq.(\ref{exp2}) with Eqs.(3,4), the identification of the
three form factors is straightforward. {\em We note that each form factor has a
real
and an imaginary part as a direct consequence of chiral symmetry}---the
explicit
$q^2$-dependence is given below.
Finally, gauge invariance of the identification of the form factors can be
shown to be satisfied \cite{Our}. 

\section{N Delta multipole transitions}

With the complete set of $N\Delta$ form factors $G_i(q^2),\, i=1,2,3$ now known
to ${\cal O}(\epsilon^3)$ one can
easily calculate \cite{Our} all $N\Delta$ multipole transitions of interest. At
present
our knowledge of the four entering couplings $b_1,b_6,D_1,E_1$ is rather poor,
but this
situation is bound to improve soon with several SSE calculations in the Delta
region
well under way. For the moment we have utilized input from the
Delta decay width and previous
phenomenological
calculations~\cite{Han} to determine the required couplings, at
$q^2=0$, and
present here
the $q^2$-evolution of the electric and coulomb multipole ratios EMR$(q^2)$,
CMR$(q^2)$
as the subsequent SSE prediction to ${\cal O}(\epsilon^3)$.

We would like to thank the organizers of Baryons 98 for giving us the
opportunity
to present these new results to the physics community.

\begin{figure}[h]
\mbox{\psfig{file=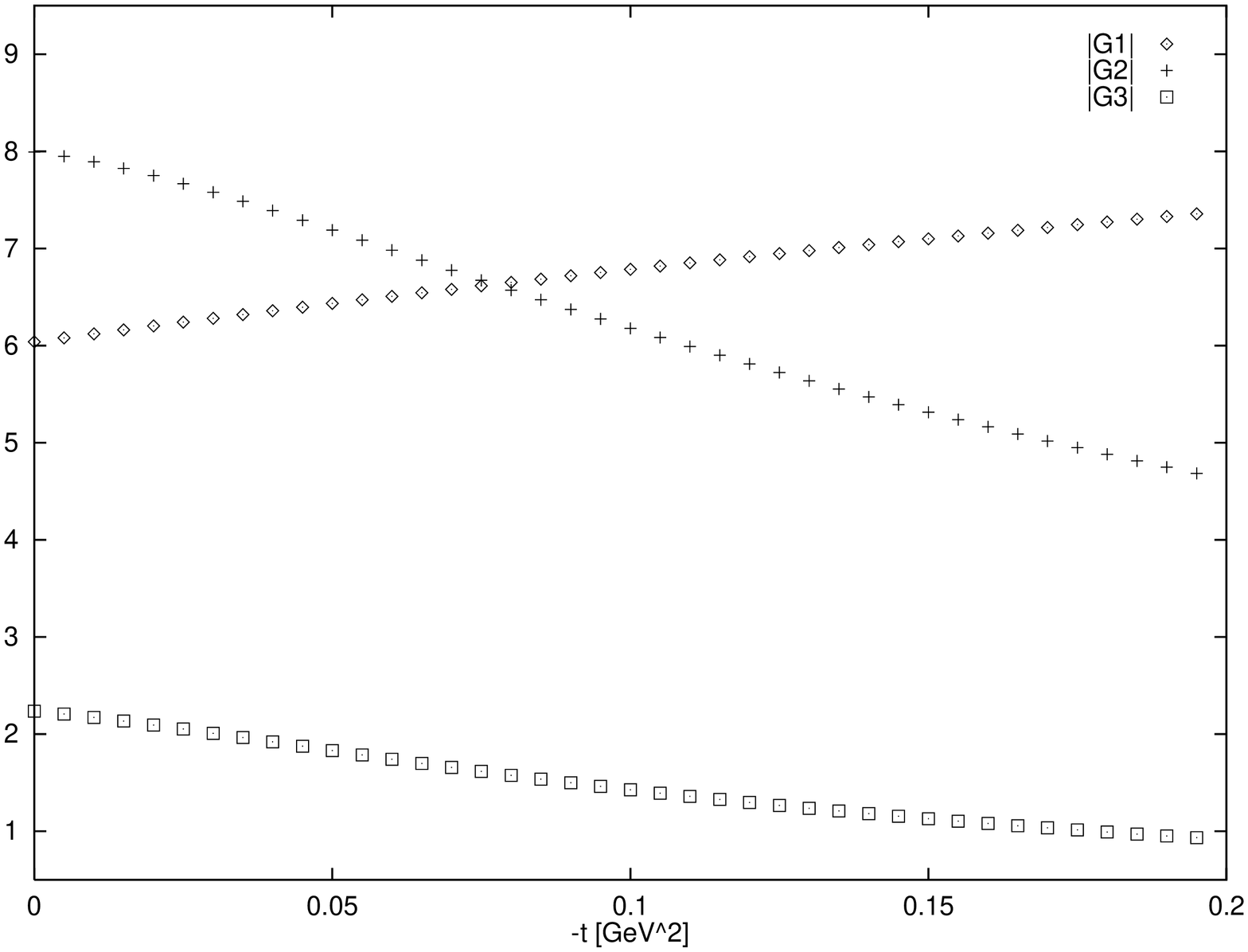,width=0.40\textwidth,angle=-90}}
\label{fig3}
\end{figure}

\begin{figure}[h]
\mbox{\epsfig{file=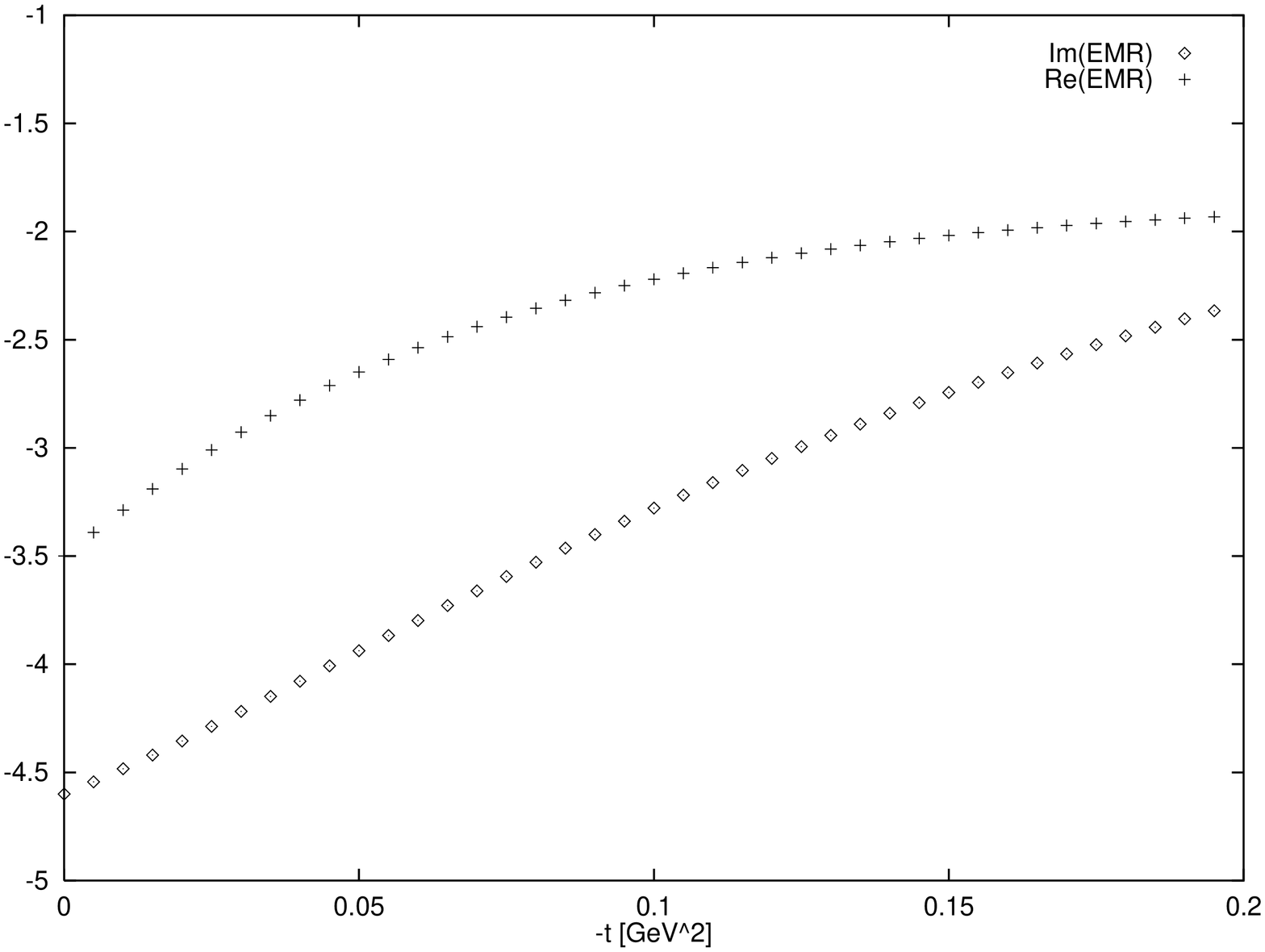,width=0.40\textwidth,angle=-90}}
\label{fig1}
\end{figure}

\begin{figure}[hb]
\mbox{\psfig{file=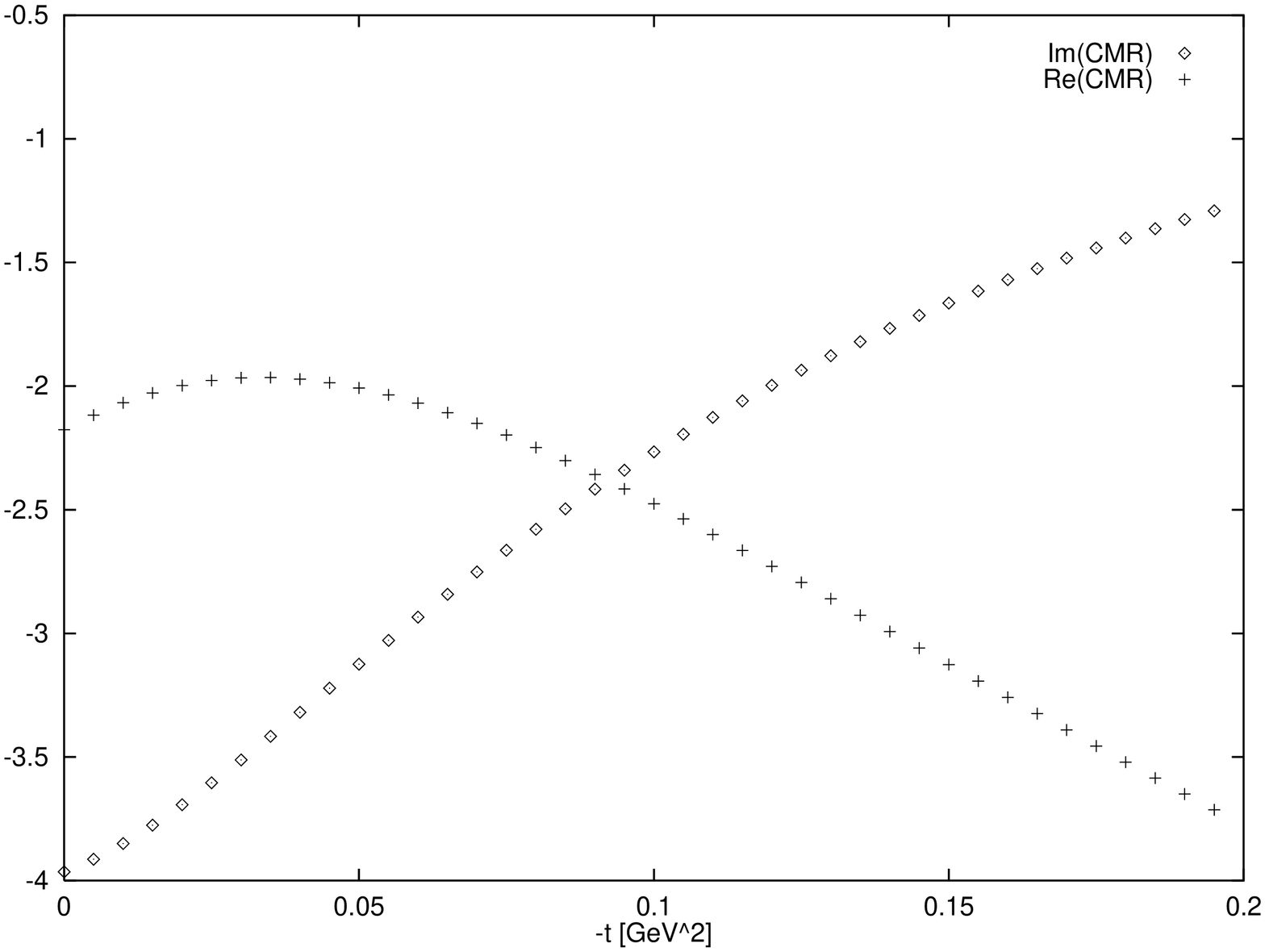,width=0.40\textwidth,angle=-90}}
\label{fig2}
\end{figure}

\end{document}